\documentclass[aps,prl,floats,twocolumn,superscriptaddress,nobibnotes]{revtex4-2}
\usepackage[svgnames]{xcolor} 
\usepackage{graphicx,epsfig}
\usepackage{times} 
\usepackage{amssymb,amsmath,multirow,rotate,color}
\usepackage{xcolor}
\usepackage[normalem]{ulem}

\usepackage{todonotes}
\usepackage[colorlinks=true,
						linkcolor=blue,
						urlcolor=blue,
						citecolor=blue,
						bookmarks=true,
						pdfborder={0 0 0}]{hyperref}

\definecolor{albicocca}{rgb}{0.98, 0.7, 0.2}
\definecolor{internationalorange}{rgb}{1.0, 0.31, 0.0}
\definecolor{giocolor}{RGB}{0, 150, 100}

\usepackage{float}
\usepackage{lipsum} 
\usepackage{ragged2e}
\usepackage{soul}
\DeclareMathAlphabet\mathbfcal{OMS}{cmsy}{b}{n}
\newcommand{\av}[1]{ \langle #1 \rangle}

\begin{document}

\title{Disentangling the Role of Heterogeneity and Hyperedge Overlap in Explosive Contagion on Higher-Order Networks}

\author{Federico Malizia}
\affiliation{Network Science Institute, Northeastern University London, London E1W 1LP, United Kingdom}
\author{Andr\'es Guzm\'an}
\affiliation{Network Science Institute, Northeastern University London, London E1W 1LP, United Kingdom}
\author{Iacopo Iacopini}
\affiliation{Network Science Institute, Northeastern University London, London E1W 1LP, United Kingdom}
\affiliation{Department of Physics, Northeastern University, Boston, MA 02115, USA}
\author{Istv\'an Z. Kiss}
\affiliation{Network Science Institute, Northeastern University London, London E1W 1LP, United Kingdom}
\affiliation{Department of Mathematics, Northeastern University, Boston, MA 02115, USA}
\date{\today}

\begin{abstract}
We introduce group-based compartmental modeling (GBCM), a mean-field framework for irreversible contagion in higher-order networks that captures structural heterogeneity and correlations across group sizes. Validated through numerical simulations, GBCM analytically disentangles the role of each interaction order to the global epidemic dynamics, revealing how heterogeneity and inter-order correlations jointly shape the onset of outbreaks and the emergence of explosive dynamics. 
Crucially, we show that inter-order correlations drive the system along distinct pathways to explosive contagion---emerging universally across both irreversible and reversible spreading processes.
\end{abstract}
\maketitle

The propagation of contagions and behaviors in complex systems often involves repeated or simultaneous stimuli that individuals receive from their social contacts~\cite{centola2007complex, hodas2014simple}. One way to mechanistically encode these non-linear effects into contagion models is to consider interactions that go beyond simple pairwise connections~\cite{battiston2020networks, torres2021and, bick2023higher}. Recent studies have highlighted the critical role of these higher-order interactions—group involving three or more individuals—in shaping the dynamics of spreading processes \cite{iacopini2019simplicial,ferraz2024contagion,ferraz2023multistability}, but also in synchronization \cite{tanaka2011multistable,millan2020explosive,skardal2020higher}, and game theory \cite{alvarez2021,civilini2021dilemmas,civilini2024explosive,guo2025evolutionary}. In fact, higher-order mechanisms~\cite{rosas2022disentangling} on complex networks give rise to a variety of new phenomena~\cite{battiston2021physics, bianconi2021higher}, such as explosive transitions~\cite{kuehn2021universal}, vanishing size of critical mass~\cite{iacopini2022group}, multi-stability~\cite{ferraz2023multistability, skardal2023multistability, zhang2024deeper} and chaos~\cite{sun2023dynamic}. Crucially, it has been shown that the way these group interactions are distributed across the system plays a central role in determining its behavior \cite{st2021universal, majhi2022dynamics, mancastroppa2023hyper, nandy2024degree, kim2024higher}. From a macroscopic point of view, the presence of hubs in higher-order structures significantly impacts the onset and evolution of spreading processes \cite{landry2020effect, st2022influential}.
More recently, the role of the microscopic arrangement of groups has been studied via two key structural concepts: {\it intra-order} correlations, describing dependencies within interactions of the same order \cite{malizia2023hyperedge}, and {\it inter-order} correlations for the interplay across orders \cite{burgio2024triadic}. While intra-order correlations depend on the microscopic arrangement of groups of the same size, inter-order correlations \cite{lamata2025hyperedge} depend on the full hierarchy of groups, and can profoundly influence collective dynamics. For instance, higher-order networks with uncorrelated sets of hyperedges behave fundamentally differently from those structured as simplicial complexes—where the downward closure requirement maximizes correlations across orders~\cite{zhang2023higher,kim2023contagion,burgio2024triadic}. 
Although significant progress has been made in characterizing the impact of structural network properties on the contagion dynamics that unfolds over it~\cite{landry2020effect, st2021universal, ferraz2021phase, st2022influential, ferraz2023multistability, kim2024higher, burgio2024triadic, malizia2023hyperedge}, capturing at the same time inter- and intra-order correlations, hyperdegree distributions, and state dependencies across different orders of interactions remains a challenge. These features introduce substantial complexity, and existing frameworks often fail to balance analytical tractability with the need to account for both structural and dynamical heterogeneity \cite{malizia2023pair,burgio2024triadic}.
\newline 
\indent In this paper, we develop a group-based mean-field framework to study the dynamics of the Susceptible-Infected-Recovered (SIR) model on higher-order networks. Our model incorporates two essential structural features—heterogeneity in hyperdegree distributions and the \textit{inter-order hyperedge overlap}~\cite{lamata2025hyperedge}, a metric which quantifies correlations between different orders of interactions. By analytically deriving the epidemic threshold and disentangling the contributions of two- and three-body interactions, we reveal how these structural features jointly shape the onset of epidemic outbreak. Crucially, we show that high heterogeneity in hyperdegree distributions can trigger explosive contagion, and that overlap modulates the early-stage activation of higher-order spreading pathways. These predictions are validated through Gillespie simulations on synthetic and empirical hypergraphs, and are shown to extend to SIS dynamics. Altogether, our results reveal the structural and dynamical mechanisms of abrupt transitions in contagion processes in the presence of group interactions, and offer a unifying framework to study spreading processes in real-world systems.
%
\newline 
\indent \textit{Modeling higher-order interactions.}---We model a system with higher-order interactions as a hypergraph $H=(\mathcal{N},\mathcal{E})$, where $\mathcal{N}$ is the set of $N=|\mathcal{N}|$ nodes that interact via $E=|\mathcal{E}|$ hyperedges, i.e., groups of nodes. Each hyperedge $e \in \mathcal{E}$, a subset of $\mathcal{N}$, can be characterized by its order $m=|e|-1$, with $m=1$ representing pairwise interactions, $m=2$ corresponding to group interactions of three nodes, etc. We call $\mathcal{E}_m$ the set of hyperedges of order $m$. Its counterpart is $k_m$, the generalized degree of order $m$, also called $k$-hyperdegree, denoting the number of $m$-hyperedges connected to a node~\cite{courtney2016generalized}. We call $P(k_m)$ their probability distribution, whose first and second moment, $\langle k_m \rangle$ and $\langle k_m^2 \rangle$, can thus be used to jointly quantify the mean connectivity and hyperdegree heterogeneity of higher-order networks. Notice, however, that these distributions alone provide no information about the microscopic arrangement of hyperedges or the correlations among different interaction orders. To quantify such correlations, we need to assess the extent to which a given structure adheres to or deviates from the inclusion property of simplicial complexes~\cite{hatcher2002algebraic}. Calling $\mathcal{F}(\mathcal{E}_{n})$ the set of $m$-cliques within $n$-hyperedges, and given two orders of interaction $m$ and $n$ ($m < n$), the \textit{inter-order hyperedge overlap} is given by~\cite{lamata2025hyperedge} 
\begin{equation}
\alpha_{m,n} = \frac{\big|\mathcal{E}_{m} \cap \mathcal{F}(\mathcal{E}_{n})\big|}{\big|\mathcal{F}(\mathcal{E}_{n})\big|},
\label{eq:inter-order-overlap}
\end{equation}
where the numerator counts the $m$-cliques within $n$-hyperedges that are also $m$-hyperedges, normalized by the total number of $m$-cliques in $n$-hyperedges. This yields $\alpha_{m,n} \in [0,1]$, with $\alpha_{m,n}=0$ indicating no overlap, and $\alpha_{m,n}=1$ maximum overlap---when all $m$-cliques in $n$-hyperedges are also $m$-hyperedges. By construction, $\alpha_{m,n} = 0$ for $m > n$. Higher-order networks constructed from empirical data exhibit a broad diversity of $\alpha_{m,n}$, as shown in SM.
\begin{figure}[t!]
	\centering
\includegraphics[width=\linewidth]{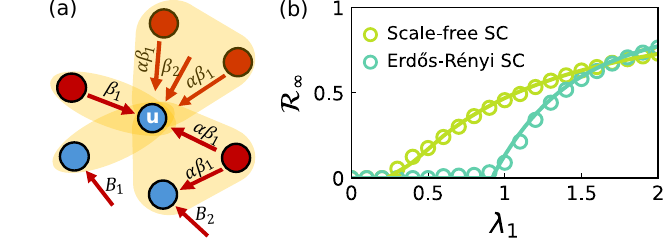}
\vspace{-2em}
	\caption{\textbf{Group-based compartmental modeling.} (a) Graphical representation of the model for $M=2$. A test node $u$ is connected to four infectious (red) and two susceptible (blue) nodes  via two 1-hyperedges and two 2-hyperedges. Arrows show the different channels of infection through the rate parameters in Eq.~\eqref{eqs:ebcm_equations_with_alpha}. (b) Final epidemic size ($\mathcal{R}_\infty$) as a function of infectivity $\lambda_1$, with $\lambda_2 = 3$. Model results (lines) are compared with Gillespie simulations (markers) on Scale-free and an Erd\H{o}s-R\'enyi simplicial complexes, respectively with $2{,}000$ and $10{,}000$ nodes (see characteristics in Table~\ref{tab:HO_networks}).} 
	\label{fig:figure_1}
    \vspace{-1em}
\end{figure}
\newline 
\indent 
\textit{Group-based compartmental modeling.---}To understand how correlations between different orders of interactions affect the onset and outcome of outbreaks, we propose a mathematical framework that explicitly includes the inter-order overlap of Eq.~\eqref{eq:inter-order-overlap} as a free parameter. To this end, we consider the edge-based compartmental modeling (GBCM) approach for the SIR process~\cite{pastor2015epidemic, kiss2017mathematics} defined in~\cite{miller2012edge,volz2011effects}, generalizing it to capture infection dynamics within groups of different orders. Each order $m=1, 2,\ldots,M$ is associated with an infection rate $\beta_m$ governing the rate at which a susceptible node is infected via a ``contagious'' $m$-hyperedge---defined as one in which all other $m$ nodes are infectious. The recovery rate for infected nodes is denoted by $\mu$, and recovered nodes cannot be reinfected.
Our formalism relies on two key quantities defined from the point of view of a test node $u$ which is part of hyperedges of different orders: $\theta_m(t)$, the probability that $u$ at time $t$ has not yet been infected by a randomly chosen infectious $m$-hyperedges it is part of; $\Phi_m^{(s,i)}(t)$, the probability that $u$ is still susceptible and member of an $m$-hyperedge containing other $s$ susceptible and $i$ infected nodes at time $t$. 
With these definitions, $\theta_m(t)$ can be expressed as $\theta_m(t) = \sum_{(s,i) \in \Omega} \Phi_m^{(s,i)}(t)$,
where $\Omega = \{(s,i) \mid 0 \leq s + i \leq m\}$ is the set of all possible combinations of $s$ susceptible and $i$ infected members of an $m$-hyperedge.
The variables $\Phi_m^{(s,i)}(t)$ are thus used to describe the progression of the epidemics within $m$-hyperedges, transitioning from fully susceptible to fully infected states. It is worth stressing that their dynamics directly depend on the inter-order overlap $\alpha_{m,n}$, which contributes to the progression of infection within hyperedges---in addition to contagion events originating from groups of different orders.
Henceforth, for simplicity, we omit the obvious time dependence. The evolution of $\theta_m$ depends on the probability that $u$ is not infected by any of the $m$-hyperedges, as given by $\dot{\theta}_m = -\beta_m \Phi_{m}^{(0,m)}$. 
Let us now leverage the formalism of probability generating function (PGFs)~\cite{newman2002spread, miller2018primer} to account for the fact that a node can take part of different hyperedges with a probability distribution $P(k_m)$. The PGF of order $m$ reads 
\begin{equation}
\label{eq:ebcm_generating_function_generalized}
    G_m(\theta_m) = \sum_{k_m=0}^\infty P(k_m)\theta_m^{k_m}.
\end{equation}
Given that, and assuming independence among interaction orders, the probability of having a given susceptible population at time $t$ is given by $\langle S \rangle = \prod_{m=1}^M \sum_{k_m} P(k_m) \theta_m^{k_m}$. This approach is particularly suited to modeling SIR dynamics, where the irreversible nature of the process allows a mapping to bond percolation on networked systems~\cite{newman2002spread, kenah2007second}. In contrast to classical SIR dynamics on dyadic networks, here we can disentangle the contribution of each interaction order to the overall epidemic. Hence, differentiating Eq.~\eqref{eq:ebcm_generating_function_generalized}, disaggregated by order and accounting for recovery, yields
\begin{equation}
\label{eq:ebcm_infecteds_generalized}
    \dot{\langle I_m \rangle} = -G'_m(\theta_m) \dot{\theta}_m - \mu \langle I_m \rangle.
\end{equation}
Finally, the total densities of infected and recovered populations at time $t$ are, respectively, $\langle I \rangle = \sum_{m=1}^M \langle I_m \rangle$ and $\langle R \rangle = 1 - \langle S \rangle - \langle I \rangle$. 
To fully appreciate and explicitly show all the components of the GBCM, we restrict our analysis %
to interactions up to order $m \leq 2$ (see Appendix for a general formulation up to any order $M$). In this case, the inter-order hyperedge overlap is simply captured by $\alpha_{1,2} \equiv \alpha$, a single parameter of interest quantifying the extent to which 2-body interactions are contained within 3-body interactions.
To further simplify the notation, we define $G(\theta_1) \equiv G_1(\theta_1)$, $H(\theta_2) \equiv G_2(\theta_2)$, and similarly $\phi_S \equiv \Phi_{1}^{(1,0)}$, $\phi_I \equiv \Phi_{1}^{(0,1)}$, $\phi_{SI} \equiv \Phi_{2}^{(1,1)}$, and $\phi_{II} \equiv \Phi_{2}^{(0,2)}$. 
\begin{figure}[t!]
	\centering
\includegraphics[width=\linewidth]{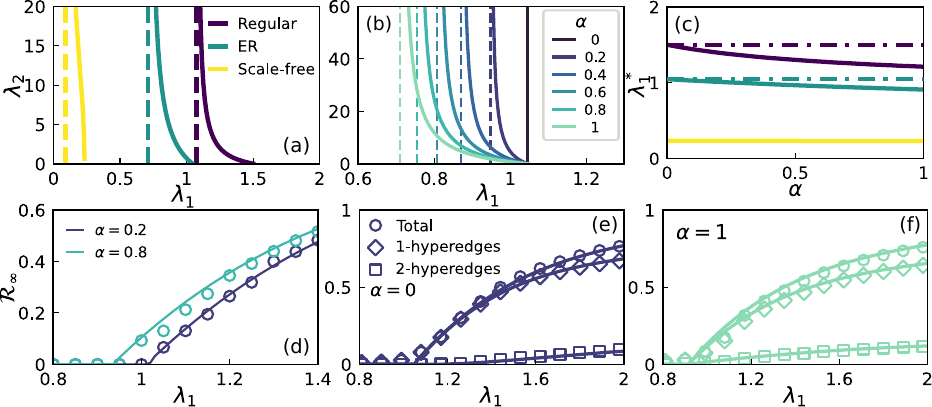}
\vspace{-2em}
	\caption{\textbf{The role of inter-order hyperedge overlap.} Epidemic thresholds in the $(\lambda_1,\lambda_2)$ plane predicted by the GBCM, Eq.~\eqref{eq:epidemic_threshold_charpoly_sm}: %
    (a) for simplicial complexes ($\alpha = 1$) in Table \ref{tab:HO_networks}; (b)  for Erd\H{o}s-R\'enyi hypergraphs with different $\alpha$ values. (c) Epidemic threshold $\lambda_1^*$ as a function of $\alpha$ (fixed $\lambda_2 = 3$) for the three classes of hypergraphs. (d-f) Comparison of the final epidemic size $\mathcal{R_\infty}$ from the GBCM model (lines) and simulations (markers) on ER hypergraphs with $N=10{,}000$ nodes, for different $\alpha$. 
    In (d) we show $\mathcal{R_\infty}$ against $\lambda_1$ for different overlap values. In (e) and (f), we consider the extreme cases $\alpha=0$ and $\alpha=1$ and show $\mathcal{R_\infty}$ disaggregated by contributions from 1- and 2-hyperedges. In all cases, $\mu = 1$.}
	\label{fig:figure_2}
\vspace{-1.5em}
\end{figure}
Under these assumptions, the resulting system of coupled equations for the GBCM approximation with $M=2$ is given by
\begin{equation}
\label{eqs:ebcm_equations_with_alpha}
    \begin{array}{ll}
    \dot{\theta_1} &= -\beta_1\phi_I; \quad \dot{\theta_2} = -\beta_2 \phi_{II},
    \\
        \dot{\phi_I} &= B_1 \phi_S -(\beta_1 + \mu)\phi_I -2\alpha\beta_1 \phi_{II},\\
\dot{\phi_{SI}}&=2B_2\phi_{SS} - (B_2 + \mu)\phi_{SI} - 2\alpha \beta_1\phi_{SI},\\
\dot{\phi_{II}}&= B_2 \phi_{SI} - (\beta_2 + 2\mu) \phi_{II} + \alpha\beta_1(\phi_{SI} - 2 \phi_{II}),
    \end{array}
\end{equation}
where $\phi_S= G'(\theta_1)H(\theta_2)/\av{k_1}$ and $\phi_{SS}= (G(\theta_1)H'(\theta_2)/\av{k_2})^2$ (see Appendix for their detailed derivation), with $\av{k_m} = \sum_{k_m} k_m P(k_m)$, corresponding to the first derivative of the PGF, defined in Eq.~\eqref{eq:ebcm_generating_function_generalized}, when $\theta_m(t) = 1$. 
Moreover, $B_1$ and $B_2$ in Eq.~\eqref{eqs:ebcm_equations_with_alpha} represent the rate of infection from external 1- and 2-hyperedges, respectively (see Appendix for their detailed expressions). 
Notice how $\alpha$ appears explicitly in the equations for the evolution of $\phi_I$, $\phi_{SI}$ and $\phi_{II}$. In particular, the term $-2\alpha \beta_1 \phi_{SI}$ accounts for the potential infections coming from pairwise interactions nested within 2-hyperedges---with similar arguments for the other terms involving $\alpha$. This formulation allows us to incorporate dynamical correlations arising from the embedding of 1- within 2-hyperedges, without system closures tailored to specific microscopic configurations.
Henceforth, we use the rescaled infectivity parameters $\lambda_1 = \av{k_1} \beta_1 / \mu$ and $\lambda_2 = \av{k_2} \beta_2 / \mu$~\cite{iacopini2019simplicial}. The ability of the formalism to capture both independent contributions from different orders and their interplay via$\alpha$ is exemplified in Fig.~\ref{fig:figure_1}(a). %
We validate our approach by comparing the final epidemic size ($\mathcal{R_\infty}$) predicted by the GBCM with averages from $500$ Gillespie simulations over different higher-order networks with $\alpha = 1$ (i.e., simplicial complexes), see Fig.~\ref{fig:figure_1}(b). The structures used to run simulations exhibit scale-free (SF) and Erd\H{o}s-R\'enyi-like (ER) hyperdegree distributions at both orders $m=1$ and $m=2$, with their characteristics summarized in Table~\ref{tab:HO_networks}. The ER and SF simplicial complexes were generated following \cite{iacopini2019simplicial} and~\cite{kovalenko2021growing}, respectively. In both cases, the GBCM predictions closely match the simulations, demonstrating the model's ability to capture the dynamics of higher-order systems.
\begin{table}[b]
    \centering
    \vspace{-1em}
    \begin{ruledtabular}
    \begin{tabular}{lcccccc}

        Higher-order networks & $\langle k_1 \rangle$ & $\av{k_1^2}$ & $\langle k_2 \rangle$ & $\av{k_2^2}$ \\
        \hline
        Regular & 6.00 & 42.00 & 1.00 & 2.00 \\
        Erd\H{o}s R\'enyi & 11.83 & 169.51 & 2.90 & 14.30 \\
        Scale-Free & 11.98 & 649.76 & 9.00 & 610.10 \\
    \end{tabular}
    \end{ruledtabular}
    \vspace{-1em}
    \caption{Characteristics of different higher-order networks considered in the study, where $\langle k_1 \rangle$ and $\langle k_2 \rangle$ denote the mean hyperdegree for pairwise and higher-order interactions, respectively, while $\av{k_1^2}$ and $\av{k_2^2}$ are the second moments of the hyperdegree distributions.}
    \label{tab:HO_networks}
    \vspace{-1em}

\end{table}
\newline 
\indent 
\textit{The role of inter-order hyperedge overlap.---}Here, we study the stability of the disease-free state of the system in Eqs.~\eqref{eqs:ebcm_equations_with_alpha} by evaluating the Jacobian at the steady state $(\theta_1, \theta_2,\phi_I, \phi_{SI}, \phi_{II})=(1,1,0,0,0)$. Despite the complexity of the model, it is possible to find an analytical expression for the epidemic threshold, revealing its explicit dependence on the interplay between structural overlap and hyperdegree heterogeneity. This, in turn, allows us to map the critical relationship between $\lambda_1$ and $\lambda_2$ at the epidemic threshold (see Appendix for the exact analytical expression).
Figure~\ref{fig:figure_2}(a) illustrates this relationship by showing the epidemic threshold in the $(\lambda_1, \lambda_2)$ plane, evaluated numerically on simplicial complexes ($\alpha=1$) with three levels of heterogeneity (as detailed in Table~\ref{tab:HO_networks}). Dashed lines indicate $\lambda_1 = \lambda_1^c$, where $\lambda_2 \to \infty$, showing that no outbreak is possible if $\lambda_1 \leq \lambda_1^c$, regardless of the value of $\lambda_2$; thus, pairwise transmission plays a dominant role. In Fig.~\ref{fig:figure_2}(b), we consider the case of ER hypergraphs with varying overlap, showing that increasing $\alpha$ consistently lowers the epidemic threshold.
To further explore the dependency of the epidemic threshold on $\alpha$, we rearrange it as a third-order polynomial in $\lambda_1$. Its solution  $\lambda_1^*$, the critical value of $\lambda_1$ at which an epidemic occurs, can be approximated using an asymptotic expansion for small $\alpha$ (see SM for details), yielding
\begin{equation}
\begin{array}{ll}
\label{eq:lambda1_asymptotic_order1}
  \lambda_1^{*} \approx & \dfrac{\av{k_1}^{2}}{\Delta_1} -  \alpha \lambda_{2} \dfrac{2 \av{k_1}^{5} \av{k_2}}{\Delta_1^{3}\left(2 \av{k_2} + \lambda_{2}\right)},
\end{array}
\end{equation}
where $\Delta_m= \Pi_m - \av{k_m}$ and $\Pi_m=\av{k_m^2}-\av{k_m} = \sum_{k_m} k_m(k_m-1) P(k_m)$. The latter expression represents the second derivative of the PGF evaluated at $\theta_m(t) = 1$. 
It is worth noting that $\Delta_m$ represents the difference between the second and first derivatives of the PGF at $\theta_m(t) = 1$.
This result highlights that stronger inter-order correlations ($\alpha > 0$) increase the system's susceptibility to outbreaks. Furthermore, when $\alpha \neq 0$, $\lambda_1^*$ depends on the strength of higher-order interactions ($\lambda_2$), in line with with recent findings \cite{burgio2024triadic,malizia2023pair}.
Additionally, Eq.~\eqref{eq:lambda1_asymptotic_order1} demonstrates that greater heterogeneity in the pairwise degree distribution (via $\Delta_1$) reduces the influence of higher-order interactions on $\lambda_1^*$.
Fig.~\ref{fig:figure_2}(c) shows the dependence of the exact epidemic threshold $\lambda_1^*$ on $\alpha$, numerically evaluated from the Jacobian matrix of the system in Eq.\eqref{eqs:ebcm_equations_with_alpha}, with $\lambda_2=3$; the dashed-dotted line denotes the $\alpha=0$ baseline. 

We next focus on the final epidemic size. In Fig.~\ref{fig:figure_2}(d-f), we compare GBCM predictions with averages from $500$ simulations on ER hypergraphs with varying $\alpha$. To create a continuous spectrum of higher-order networks with overlap $\alpha$ spanning the full range from 1 to 0, we rewired the layer of 1-hyperedges of an ER simplicial complex while preserving its 1-hyperdegree distribution and the structure of the 2-hyperedges (see SM for details on network generation and rewiring).
Figure\ref{fig:figure_2}(d) confirms that the GBCM accurately predicts the epidemic threshold, which depends on the level of inter-order overlap in the underlying higher-order networks: when $\alpha$ increases, the epidemic starts earlier. Additionally, in Fig.\ref{fig:figure_2}(e)-(f) we explicitly separate the contributions coming from pairwise and higher-order contagions to the final epidemic size for $\alpha=0$ and $\alpha=1$, respectively. For $\alpha=0$, three-body transmission is delayed, requiring a buildup of infected nodes to activate higher-order contagion. In contrast, for $\alpha=1$, both pairwise and higher-order contagion modes activate simultaneously at $\lambda_1^*$, as predicted by Eq.~\eqref{eq:epidemic_threshold_charpoly_sm}. 
%
\newline 
\indent 
\textit{High heterogeneity of group interactions leads to explosive phenomena.---}We now systematically analyze how heterogeneity in higher-order interactions (variations in hyperdegree distribution) affect epidemic dynamics, focusing on both the final epidemic size and the temporal evolution of outbreaks.
\begin{figure}[t!]
\centering\includegraphics[width=\linewidth]{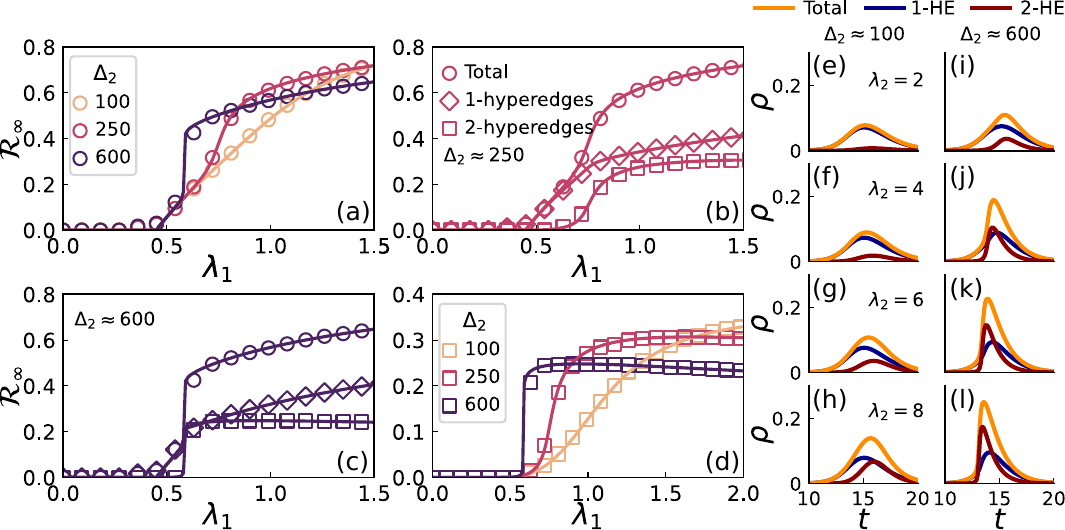}
\vspace{-2em}
\caption{\textbf{High heterogeneity of group interactions leads to explosive phenomena.} 
(a) Final epidemic size ($\mathcal{R}_\infty$) from GBCM predictions (solid lines) and Gillespie simulations (circles), showing excellent agreement. (b-d) Illustration of the double-transition process: (b-c) compare $\mathcal{R}_\infty$ for different heterogeneity levels in 2-hyperedges, where (c) highlights continuous transitions at $\lambda_1^*$ (pairwise interactions) and abrupt jumps at $\hat{\lambda}_1$ (high heterogeneity). (d) Contribution of 2-hyperedges to $\mathcal{R}_\infty$ for increasing $\Delta_2$, revealing sharper transitions with increasing heterogeneity. (e-l) Temporal evolution of the total prevalence $\rho$ (orange) and contributions from pairwise (blue, 1-HE) and three-body (red, 2-HE) interactions with varying $\lambda_2$. Curves are obtained via the GBCM for hypergraphs featuring negative binomial hyperdegree distributions ($\alpha=0$) with $\Delta_2 \approx 100$ (e-h) and $\Delta_2 \approx 600$ (i-l).}\label{fig:figure_3}
\vspace{-2em}
\end{figure}
We consider three synthetic hypergraphs, each composed by $N = 10{,}000$ nodes, whose pairwise and three-body interactions follow  uncorrelated negative binomial hyperdegree distributions~\cite{kiss2023necessary}. This allows to independently tune the variance of the $2$-hyperdegree distribution while keeping the mean degrees fixed. The mean pairwise degree is $\langle k_1 \rangle \approx 12$ with $\av{k_1^2} = 327$. For three-body interactions, we consider three levels of heterogeneity with $\langle k_2 \rangle \approx 9$ , quantified by $\Delta_2 \approx [100, 250, 600]$, namely the difference between the second and first derivatives of the PGF at $\theta_2(t) = 1$ (see SM for details).
Since hyperdegree distributions are uncorrelated, inter-order overlap is zero ($\alpha = 0$) in all cases.
Figure~\ref{fig:figure_3}(a) shows the final epidemic size as a function of $\lambda_1$ for $\lambda_2 = 6$ $(\beta_2 \approx 0.66$) across three levels of heterogeneity in $P(k_2)$. Comparing GBCM predictions against averages over $500$ simulations, we again show that the GBCM accurately captures the system's behavior across all heterogeneity levels. Notably, as $\Delta_2$ increases, the epidemic transitions in $\mathcal{R}_{\infty}$ become more abrupt.
Figures~\ref{fig:figure_3}(b)-(c) further decompose the contributions from $1$- and $2$-hyperedges to the final epidemic size for 
$\Delta_2 \approx 250$ and $\Delta_2 \approx 600$, respectively. In both cases, a double transition occurs due to $\alpha = 0$. At $\lambda_1^*$, corresponding to the epidemic threshold derived from Eq.~\eqref{eq:epidemic_threshold_charpoly_sm}, the system exhibits a continuous transition primarily driven by \(1\)-hyperedges. However, as shown in Fig.~\ref{fig:figure_3}(b), a secondary increase in $\mathcal{R}_{\infty}$ appears at $\hat{\lambda}_1$, marking the onset of higher-order contagion.
In contrast, at higher heterogeneity levels ($\Delta_2 \approx 600$), Fig.~\ref{fig:figure_3}(c) reveals hybrid transitions for 
dominated by abrupt contributions from $2$-hyperedges. Finally, Fig.~\ref{fig:figure_3}(d) highlights how increasing $\Delta_2$ sharpens the transition in $\mathcal{R}_{\infty}$, confirming the role of heterogeneity in driving explosive phenomena, even in the absence of inter-order overlap. Further evidence on the emergence of explosive dynamics in SIR processes on empirical higher-order networks is reported in the SM.

To better understand the mechanism leading to explosive contagion, we examine the temporal evolution of the epidemic.  Fixing the pairwise infectivity at $\lambda_1 = 1$ and varying $\lambda_2$, we track the prevalence $\rho$ over time, again decomposing the contributions to the infection from different orders via the GBCM.
Figure~\ref{fig:figure_3}(e-h) show that for low heterogeneity ($\Delta_2 \approx 100$), infections via 2-hyperedges are delayed until parwise interactions generate a critical mass of infectious nodes. For high heterogeneity ($\Delta_2 \approx 600$), shown in Fig.~\ref{fig:figure_3}(i-l), higher-order contagion rapidly amplifies epidemics, with 2-hyperedges driving the explosive growth once pairwise spreading passes a threshold (k,l).

An intuitive explanation for explosive contagion can be gained by considering the density of infected individuals through 2-hyperedges, given by $\dot{\langle I_2 \rangle} = -H'(\theta_2) \dot{\theta}_2 - \mu \langle I_2 \rangle$.
Since explosive behavior occures when $\dot{\langle I_2 \rangle} \rightarrow \infty$ at some time $\hat{t}>0$, the condition for the critical value $\hat{\lambda}_2$ reads 
\begin{equation}
\label{eq:explosion_condition}
    \hat{\lambda}_2 \approx \frac{2 \langle k_2 \rangle^2 \left(\langle k_1 \rangle + \alpha \lambda_1 \right)}{\langle k_1 \rangle \left(\Delta_2 \hat{\phi}_{SI} + \langle k_2 \rangle (\hat{\phi}_{SI} - 1) \right)},
\end{equation}
where $\hat{\phi}_{SI}$ represents the critical density of $\phi_{SI}$ required to activate contagion via 2-hyperedges, constrained by $\epsilon \leq \hat{\phi}_{SI} \leq 1$ (see SM for the full derivation). 
This result shows that higher values of $\Delta_2$ reduce the critical $\hat{\lambda}_2$, making explosive behavior more likely, whereas lower $\Delta_2$ suppress such phenomena by increasing $\hat{\lambda}_2$. Despite the lack of an exact analytical expression for $\hat{\phi}_{SI}$, Eq.~\eqref{eq:explosion_condition} provides an important explanation of the role of heterogeneity in the three-body interactions, both from a structural and dynamical viewpoint.

\begin{figure}[t]
\centering\includegraphics[width=1\linewidth]{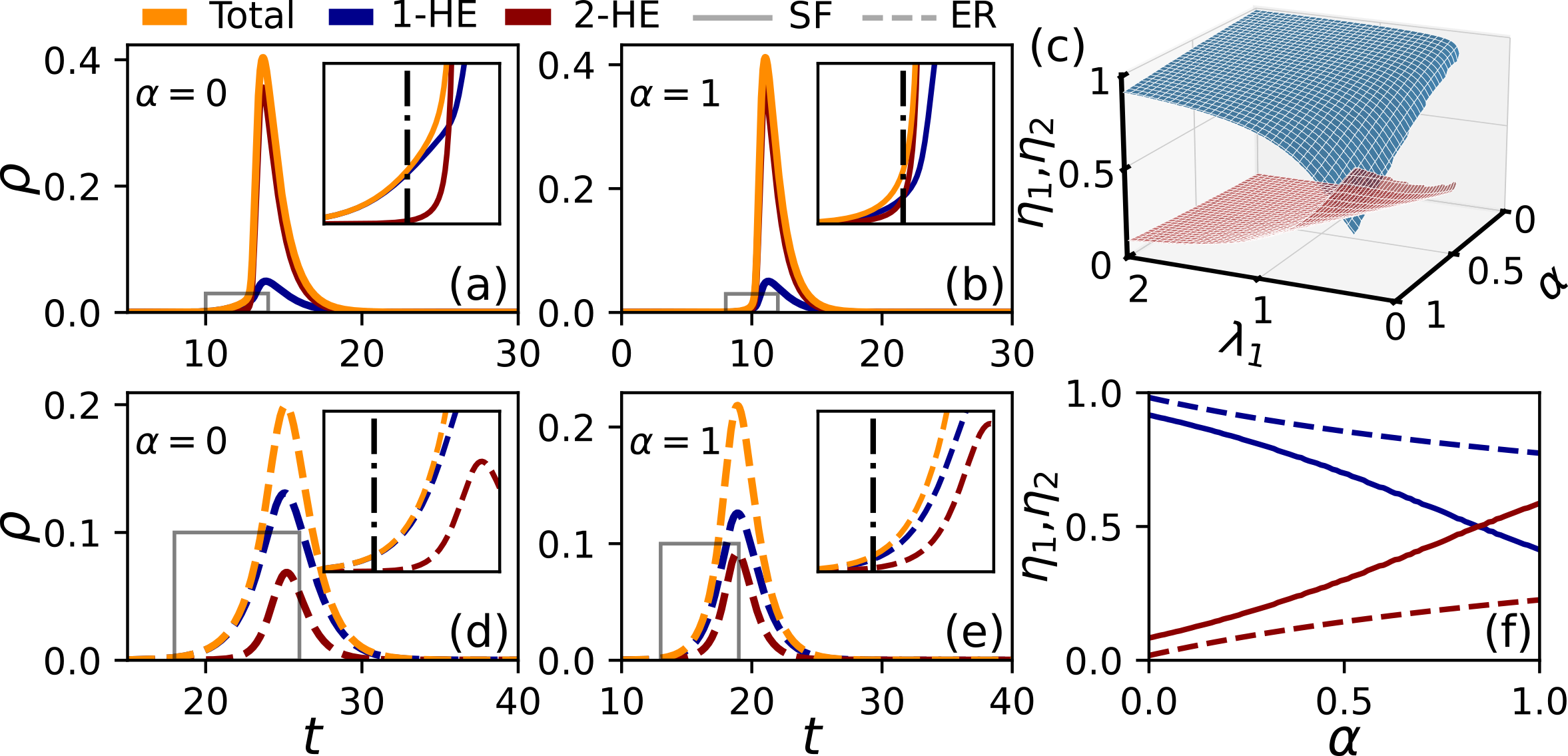}\vspace{-1em}
\caption{\textbf{Inter-order hyperedge overlap modulates early-stage contributions to explosive contagion.} 
Temporal evolution of prevalence from GBCM for Scale-Free (a-b) and Erd\H{o}s-Rényi (d-e) hypergraphs (see Table~I), for $\lambda_2 = 15$, with $\alpha = 0$ and $\alpha = 1$. Values of $\lambda_1$ are chosen to yield $\mathcal{R}_\infty = 0.8$ in each case. Vertical lines mark the time $t^\dagger$ at which the total $\rho(t^\dagger) = 0.01$. (c) early-stage contributions $\eta_1$ (blue) and $\eta_2$ (red) evaluated at $t^\dagger$ across the $(\lambda_1, \alpha)$ space for SF hypergraphs. The same in (f) as a function of $\alpha$ for SF (solid lines, $\lambda_1 = 0.4$) and ER (dashed lines, $\lambda_1 = 1.6$), both at $\lambda_2 = 15$.}
\label{fig:figure_4}
\vspace{-2em}
 \end{figure}

\textit{Inter-order hyperedge overlap modulates early-stage contributions to explosive contagion.---} Having highlighted the role of hyperdegree heterogeneity in triggering abrupt epidemic surges, we now conclude our analysis by investigating how the inter-order hyperedge overlap shapes the early-stage dynamics that lead to explosive behavior.
To this end, we consider the SF and ER hypergraphs in Table~\ref{tab:HO_networks} which allow us to explore the full range of $\alpha$ values under two different levels of heterogeneity.
Figures~\ref{fig:figure_4}(a)-(b) and (d)-(e) show the temporal evolution of infected densities predicted by the GBCM for the SF and ER cases, respectively, at the extremes $\alpha = 0$ and $\alpha = 1$.
As expected, high heterogeneity leads to explosive dynamics in both network classes. However, the effect of the overlap is evident: for $\alpha=0$, the activation of 2-hyperedge contagion is delayed, requiring a critical density of pairwise infections to trigger group-level transmission. In contrast, for $\alpha = 1$, the pairwise and higher-order components emerge simultaneously. This suggests that overlap acts as a catalyst, lowering the barrier for early activation of higher-order contagion.
This behavior can be linked to the divergence $\dot{\langle I_2 \rangle} \to \infty$ at $t = \hat{t}$, which marks the explosive onset driven by higher-order contagion. For $t < \hat{t}$, the spread is predominantly pairwise-driven; at $t \geq \hat{t}$, contagion rapidly transitions to being higher-order dominated.
This transition marks a tipping point in the spreading dynamics. To quantify it empirically, we define the early-stage contributions $\eta_m = \bar{\rho}_m(t^\dagger)/\sum_m \bar{\rho}_m(t^\dagger)$,
where $\bar{\rho}_m(t)$ denotes the cumulative infection density via $m$-hyperedges, ignoring recovery, and evolves as $\dot{\bar{\rho}}_m = -G'_m(\theta_m)\dot{\theta}_m$. We set $t^\dagger$ as the time when the total prevalence reaches $\rho(t^\dagger) = 0.01$, representing the onset of macroscopic contagion.
Although $t^\dagger$ does not coincide exactly with the singularity at $\hat{t}$, $\eta_m$ offers a practical and model-grounded proxy for the interaction order seeding the outbreak.
Figure~\ref{fig:figure_4}(c) shows the early-stage contributions $\eta_1$ (blue) and $\eta_2$ (red) across the $(\lambda_1, \alpha)$ plane for SF hypergraphs ($\lambda_2=15$). While pairwise infections dominate most of the space, for high overlap and low $\lambda_1$ (near the epidemic threshold) three-body interactions begin to contribute significantly, and even dominate.
Importantly, while the mechanism of activation differs across levels of overlap, all regimes exhibit explosive transitions. Hence, overlap reshapes---but does not suppress---the emergence of explosive contagion.
Figure~\ref{fig:figure_4}(f) summarizes $\eta_1$ and $\eta_2$ as a function of $\alpha$ for both SF (solid) and ER (dashed) structures. In the ER with low heterogeneity, increasing $\alpha$ directly enhances early infections via 2-hyperedges, given the large $\lambda_2$. In the SF case, the influence of overlap is more nuanced: at high $\alpha$, 1- and 2-hyperedges contribute comparably, whereas at low $\alpha$, pairwise infections dominate initially and group transmission only ignites later.
Taken together, these findings clarify the distinct role of hyperedge overlap in shaping the microscopic routes to spread, by regulating when higher-order contagion activates and how it contributes to explosive contagion. In SM, we further extend this analysis across a broader parameter space and confirm these effects via Gillespie simulations, including for SIS dynamics.
\newline 
\indent 
\textit{Conclusions.---}In this paper, we introduced a group-based mean-field framework for irreversible contagion processes on higher-order networks, which incorporates both heterogeneity in hyperdegree distributions and inter-order correlations. Applied to two- and three-body interactions, our approach accurately predicts epidemic thresholds, disentangles the contributions of each order of interaction, and explains the emergence of abrupt, explosive transitions driven by strong higher-order infectivity and group heterogeneity.
By combining analytical results and extensive Gillespie simulations, we demonstrated how inter-order overlap modulates the onset of higher-order contagion, thereby shaping early-stage dynamics to explosive contagion.
While we focused on SIR dynamics for analytical tractability, our findings persist in reversible SIS processes, confirming the generality of our results.
However, extending the GBCM formalism to reversible dynamics remains an open challenge, as the breakdown of the percolation mapping may require approximations that compromise accuracy and generality.\newline
Our approach stresses the importance of higher-order network features in shaping the contagion dynamics and provides a foundation for the development of more advanced frameworks that capture key structural features of real-world systems. This could be further extended to account for their dynamic~\cite{burgio2023adaptive} and temporal~\cite{gallo2024higher, iacopini2024temporal} nature, or to access their impact on multiple interacting processes~\cite{min2018competing, lucas2023simplicially, andres2024competition}.

\textit{Acknowledgments.---} Authors acknowledge Joel C. Miller for valuable comments.

\textit{Appendix A: Group-based approximation modeling up to order $M$.---}
We consider a Susceptible-Infected-Recovered (SIR) process with higher-order interactions of size $m = 1, \dots, M$. Each order $m$ has an associated infection rate $\beta_m$, which represents the rate at which a susceptible node becomes infected when connected to an $m$-hyperedge where all $m$ neighbors are infected. The recovery rate is given by $\mu$.

To describe the infection dynamics in a general form, we define $\theta_m(t)$ as the probability that, at time $t$, a test node $u$ has not been infected through a randomly chosen $m$-hyperedge from those it belongs to. If the test node $u$ is connected to $k_m$ distinct $m$-hyperedges, then the probability that $u$ has not been infected through any of them by time $t$ is given by $\theta_m(t)^{k_m}$. Using the probability generating function (PGF) $G_m(x)$ of the $m$-hyperedge degree distribution, the probability that a randomly chosen node has not received the disease via any $m$-hyperedge is given by $G_m(\theta_m(t)) = \sum_{k_m=0}^\infty P(k_m) \theta_m(t)^{k_m}$. The probability of a node being susceptible at time $t$ is the product of these probabilities across all orders of interaction, which corresponds to the average susceptible population:

\begin{equation}
\label{eq:ebcm_susceptible_generalized}
    \langle S(t) \rangle = \prod_{m=1}^M G_m(\theta_m(t)) = \prod_{m=1}^M \sum_{k_m=0}^\infty P_m(k_m) \theta_m(t)^{k_m}.
\end{equation}

If a test node $u$ is in an $m$-hyperedge containing $s$ susceptible, $i$ infected, and $m - (s + i)$ recovered neighbors, the probability that $u$ remains uninfected through this $m$-hyperedge is defined as $\Phi_m^{(s,i)}(t)$. Thus, $\theta_m(t)$ can be decomposed as:
\begin{equation}
    \theta_m(t) = \sum_{(s,i) \in \Omega} \Phi_m^{(s,i)}(t),
\end{equation}
where $\Omega = \{(s,i) \mid 0 \leq s+i \leq m\}$. For example, $\Phi_5^{(1,3)}(t)$ represents the probability that a test node is in a $5$-hyperedge with $1$ susceptible, $3$ infected, and $1$ recovered neighbor, and has not been infected up to time $t$. 

The temporal evolution of $\theta_m(t)$ is governed by:
\begin{equation}
    \dot{\theta}_m(t) = -\beta_m \Phi_m^{(0,m)}(t),
\end{equation}
since infections through an $m$-hyperedge occur only when all its $m$ neighbors are infected. The evolution of $\Phi_m^{(0,m)}(t)$ depends on transitions between all $\Phi_m^{(s,i)}(t)$ states in $\Omega$. These transitions are influenced by external infections, internal infections, and recoveries, leading to recursive dependencies. The recursive dependencies lead to the variable $\Phi_{m}^{(m,0)}$, representing the probability that a test node $u$ has not been infected via an $m$-hyperedge because all its members remain susceptible. To compute $\Phi_{m}^{(m,0)}$, consider a neighbor $v$ of node $u$ within an $m$-hyperedge. The degree of $v$ follows the excess distribution $Q_m(k_m) = k_m P(k_m)/\langle k_m \rangle$. The probability that node $v$ has not been infected via any of its other $m$-hyperedges is $\theta_m^{k_m - 1}$, and that it remains susceptible through interactions of any other order $n$ is $\theta_n^{k_n}$, where $Q_n(k_n) = P_n(k_n)$. Combining these factors, the probability that all $m$ nodes in an $m$-hyperedge to which $u$ belongs are susceptible is:
\begin{equation}
    \Phi_{m}^{(m,0)} = \left(\frac{G_m'(\theta_m)}{\langle k_m \rangle} \prod_{n \neq m} G_n(\theta_n)\right)^{m}.
    \label{eq:phi_m_general}
\end{equation}

At the start of the epidemic, $\theta_m(t) \approx \Phi_m^{(m,0)}(t) \approx 1$ for any $m$. Transitions from $\Phi_m^{(m,0)}$ to $\Phi_m^{(m-1,1)}$ occur due to external infections, where a susceptible neighbor of $u$ becomes infected through another group. The rate of external infections for an $m$-hyperedge, denoted $B_m$, can be expressed as:
\begin{widetext}
\begin{equation}
\label{eq:outside_infecive_rate}
    B_m = - \frac{G_m''(\theta_m) \prod_{n \neq m} G_n(\theta_n) \dot{\theta}_m + \sum_{n \neq m} G_m'(\theta_m) G_n'(\theta_n) \prod_{p \neq m,n} G_p(\theta_p) \dot{\theta}_n}{G'_m(\theta_m) \prod_{n \neq m} G_n(\theta_n)}.
\end{equation}
\end{widetext}

 The probabilities $\Phi_m^{(s,i)}(t)$ are treated as compartments, and the transitions between them are described by a set of differential equations. The rate of change $\dot{\Phi}_m^{(s,i)}$ accounts for eight terms: five decreasing (external infection of susceptible neighbors, internal infection of susceptible neighbors from lower orders and higher orders, internal infection of test node and recovery) and three increasing (infection of a susceptible neighbor or the test node itself, and recovery of an infected neighbor).

Considering these transitions, the differential equation for $\dot{\Phi}_m^{(s,i)}$ is:
\begin{widetext}
    \begin{equation}
    \begin{split}
        \dot{\Phi}_m^{(s,i)} = &- sB_m \Phi_m^{(s,i)}(t) - \sum_{j=1}^i s \binom{i}{j} \alpha_{j,m}\beta_j \Phi_m^{(s,i)}(t) - \delta_{i,0}^{*}\sum_{j=1}^i \binom{i}{j} \alpha_{j,m}\beta_j \Phi_m^{(s,i)}(t) \\ &  - \delta_{i,m}\delta_{M,m}^*\sum_{k=m+1}^{M} \sum_{j=1}^{k-1} (k-j+1) \alpha_{j,k} \beta_j \Phi_k^{(0,k)}  - \mu i \Phi_m^{(s,i)}(t) 
         + \delta_{s+i,m}^{*} (s+1)B_m \Phi_m^{(s+1,i-1)} \\ & + \delta_{s+i,m}^{*} \mu (i+1) \Phi_m^{(s,i+1)} + \delta_{i,0}^{*}\sum_{j=1}^{i-1}(s+1) \binom{i-1}{j} \alpha_{jm} \beta_j \Phi_m^{(s+1,i-1)},
    \end{split}
\end{equation}
\end{widetext}

where $\alpha_{i,j}$ is the inter-order overlap between $i$- and $j$-hyperedges, $\delta_{i,j}$ represents the Kronecker delta and $\delta_{i,j}^{*}=(1-\delta_{i,j})$. By leveraging $\alpha_{i,j}$, we are able to to incorporate dynamical correlations arising from the embedding of i- within j-hyperedges, without system closures tailored to specific microscopic configurations.

The system is solved numerically using the following equations:
\begin{equation}
    \begin{split}
        \dot{\boldsymbol{\Theta}} &= -\boldsymbol{\beta} \boldsymbol{\Phi}_I, \\
        \dot{\boldsymbol{\Phi}}_1 &= \boldsymbol{f}(\boldsymbol{\beta}, \boldsymbol{B}, \boldsymbol{\Phi}_1,\boldsymbol{\Phi}_2...\boldsymbol{\Phi}_M), \\
        \vdots & \\
        \dot{\boldsymbol{\Phi}}_M &= \boldsymbol{f}(\boldsymbol{\beta}, \boldsymbol{B}, \boldsymbol{\Phi}_1,\boldsymbol{\Phi}_2...\boldsymbol{\Phi}_M),
    \end{split}
\end{equation}
where we used the following notations $\boldsymbol{\beta}=\{\beta_1, \beta_2... \beta_M\}$, $\boldsymbol{B}=\{B_1, B_2... B_M\}$, $\boldsymbol{\Theta}=\{\theta_1, \theta_2... \theta_M\}$,$\boldsymbol{\Phi}_I=\{\Phi_1^{(0,1)}, \Phi_2^{(0,2)}... \Phi_M^{(0,M)}\}$  and  $\boldsymbol{\Phi}_m$ represents the set of $\Phi_m^{(s,i)}(t)$ such that $(s,i) \in \Omega$.

\textit{Appendix B: Epidemic threshold for $M=2$.---} To derive the epidemic threshold for the GBCM in the case of $M=2$, we consider the system in Eqs.\eqref{eqs:ebcm_equations_with_alpha}. 
Here, $B_1$ and $B_2$ represent the rates of infection to a susceptible node connected to the test node $u$ through 1-hyperedges and 2-hyperedges, respectively. From Eq.~\eqref{eq:outside_infecive_rate}, they are defined as:
\begin{equation}
\label{eq:outside_infecive_rate_m2}
    \begin{array}{ll}
        B_1 &= - \dfrac{G''(\theta_1) H(\theta_2)\dot{\theta}_1 + H'(\theta_2) G'(\theta_1) \dot{\theta}_2}{G'(\theta_1) H(\theta_2)}, \\[10pt]
        B_2 &= - \dfrac{G'(\theta_1) H'(\theta_2)\dot{\theta}_1 + G(\theta_1) H''(\theta_2) \dot{\theta}_2}{G(\theta_1) H'(\theta_2)}.
    \end{array}
\end{equation}

Furthermore, given the definition in Eq.\eqref{eq:phi_m_general}, we obtain $\Phi_{1}^{(1,0)} = \phi_S = G'(\theta_1)H(\theta_2)/\langle k_1 \rangle$ and $\Phi_{2}^{(2,0)} = \phi_{SS} = (G(\theta_1)H'(\theta_2)/\langle k_2 \rangle)^2$. Given that, to assess the stability of the system, we substitute the equations for $\dot{\theta_1}$ and $\dot{\theta_2}$ given by Eqs.~\eqref{eqs:ebcm_equations_with_alpha} and evaluate the Jacobian matrix of the system 
around the disease-free equilibrium $(\theta_1, \theta_2, \phi_I, \phi_{SI}, \phi_{II}) = (1,1,0, 0, 0)$. By considering the free term in the characteristic polynomial of the Jacobian matrix, we derive the epidemic threshold as
\begin{widetext}
\begin{equation}
\label{eq:epidemic_threshold_charpoly_sm}
\lambda_2=\dfrac{2 \av{k_2}^2 \left[\left( \av{k_1} + \alpha \lambda_1\right) \left( \av{k_1} + 2 \alpha \lambda_1 \right) \left( \av{k_1}^2 - \lambda_1 \Delta_1 \right) - \Pi_1(\av{k_1}-1)\av{k_1}^2\lambda_1 + 2\alpha^2\av{k_1}^4\lambda_1^3\right] }{\av{k_1} \left[\av{k_1}^3 \av{k_2} -\av{k_1}\av{k_2} \Delta_1\lambda_1 - 2 \av{k_1}^2 \Delta_2\alpha \lambda_1 + 2\left[ \Omega_{1,2} - 2\left(\av{k_1} \Pi_2 + \av{k_2}\Pi_1\right) \right] \alpha \lambda_1^2\right]},
\end{equation}
\end{widetext}
where $\Pi_m=\av{k_m^2}-\av{k_m} = \sum_{k_m} k_m(k_m-1) P(k_m)$ represents the second derivative of the PGF evaluated at $\theta_m(t) = 1$. 
Additionally, $\Delta_m= \Pi_m - \av{k_m}$ captures the difference between the second and first derivatives of the PGF at $\theta_m(t) = 1$, 
and $\Omega_{m,n} = \av{k_m^2}\av{k_n^2} - \av{k_m}^2 \av{k_n}^2$.
The detailed calculations are provided in the Supplementary Material (SM).

\bibliographystyle{unsrt}
\bibliography{biblio.bib}
\end{document}